\documentstyle[12pt]{article}

\def\beq{\begin{equation}}
\def\eeq{\end{equation}}

\def\bea{\begin{eqnarray}}
\def\eea{\end{eqnarray}}
\def\ben{\begin{enumerate}}
\def\een{\end{enumerate}}

\def\nn{\nonumber}

\def\brr{\begin{array}}
\def\err{\end{array}}

\begin{document}

\hfill UB-ECM-PF 95/

\hfill April 1995


\begin{center}

{\LARGE \bf
A renormalization group improved non-local gravitational
effective Lagrangian}

\vspace{1cm}

{\sc E. Elizalde}\footnote{E-mail: eli@zeta.ecm.ub.es} \\
Center for Advanced Studies CEAB, CSIC, Cam\'{\i} de Santa
B\`arbara,
17300 Blanes,
\\ and Department ECM and IFAE, Faculty of Physics,
University of Barcelona, \\ Diagonal 647, 08028 Barcelona,
Catalonia, Spain \\
 and \\
{\sc S.D. Odintsov}\footnote{E-mail: odintsov@ecm.ub.es.
On
leave from: Tomsk Pedagogical Institute, 634041 Tomsk, Russia.}
\\
Department ECM, Faculty of Physics,
University of  Barcelona, \\  Diagonal 647, 08028 Barcelona,
Catalonia,
Spain \\

\vspace{2cm}

{\bf Abstract}
\end{center}

Renormalization group techniques are used in order to obtain the
improved non-local gravitational effective action corresponding to any
asymptotically
free GUT, up to invariants which are quadratic on the curvature. The
corresponding
non-local gravitational equations are written down, both for the
case of
asymptotically free GUTs and also for quantum R$^2$-gravity.
The implications of the results when obtaining the flux of
vacuum radiation through the future null infinity are briefly
discussed.

\vfill

\newpage


The effective action has turned out to be a quite important
subject in
the study of different aspects of quantum field theory. Among
the
phenomena to which it has been applied successfully,  we can
mention
symmetry breaking/restoration effects, phase transitions in
general,
models of quantum corrected field equations, etc.

Most of the studies of the effective action have been limited to
a
quasi-local approach (for a general introduction see \cite{8,1}),
that
is, they deal with almost constant background fields, as is the
case of
quantum gravity on a De Sitter background \cite{18} ---which is
important in inflationary universes. Recently, some interest has
arisen
(\cite{9}-\cite{12},
see \cite{11} for an extense account) for the case of weak but
very
quickly varying background fields, which typically lead to
non-local
effective actions. In the present note, by using simple
renormalization
group (RG) methods ---implemented by means of a Wilsonian
procedure \cite{5}--- we are going to show how one can obtain in
fact an
improved non-local effective gravitational action for a big class
of
theories.

The starting point for our considerations will be a massless,
multiplicatively renormalizable theory including  scalar, spinor
and
vector fields on a classical gravitational background. The
corresponding Euclidean Lagrangian has the following form
\bea
L &=& L_m + L_{ext}, \nn \\
L_m &=& L_{YM} + \frac{1}{2}
(\nabla_\mu \varphi )^2 +
\frac{1}{2} \xi R \varphi^2
 + \frac{1}{4!} f \varphi^4
+i\overline{\psi} (\gamma^\mu \nabla_\mu-h \varphi ) \psi, \nn \\
L_{ext} &=& a_1R^2 + a_2 C^2_{\mu\nu\alpha\beta} +a_3 G + a_4
\Box R.
\label{541}
\eea
By choosing a specific gauge group, we can assume that some
multiplets
of the scalar, $\varphi$, and spinor, $\psi$, fields are given in
some
concrete representation of the gauge group.

We will assume that our theory (\ref{541}) is a typical
asymptotically
free GUT in curved spacetime (for a general introduction, see
\cite{1}).
In principle one could equally well consider other types of GUTs,
what
would not change qualitatively the conclusions of our study
below.

The running coupling constants corresponding to the
asymptotically free
couplings of the theory (\ref{541}) have the form \cite{3,4}
\bea
&& g^2(t) =g^2 \left[ 1 + \frac{B^2g^2t}{(4\pi)^2} \right]^{-1},
\ \ \ \ g^2(0)=g^2, \nn \\
&& h^2(t) =\kappa_1g^2(t) , \ \ \ \ \ \
f(t) =\kappa_2g^2(t),
\label{542}
\eea
where $t$ is the RG parameter while $\kappa_1$ and  $\kappa_2$
are
numerical couplings
defined by the specific features of the theory under
consideration. We
know of many examples of such theories, with gauge groups SU(N),
O(N),
E$_6$, etc. \cite{3,4}.
Asymptotic freedom ($g^2(t) \rightarrow 0$, $t \rightarrow
\infty$) is
realized for all running couplings: gauge,
Yukawa and scalar ones, as is easy to see from (\ref{542}).

The study of asymptotically free GUTs in curved
spacetime was started in Ref. \cite{2} (for a review and detailed
list
of references see \cite{1}). In the theories with one scalar
multiplet,
for the running scalar-graviton coupling constant one gets
\beq
\xi (t) =\frac{1}{6} + \left( \xi -  \frac{1}{6} \right) \left[ 1
+
\frac{B^2g^2t}{(4\pi)^2} \right]^b,
\label{543}
\eeq
where $\xi (0) =\xi$ and where for the different
GUTs the constant $b$ can be either positive,
negative or zero (see Ref. \cite{1}).

The gravitational running couplings are defined by the following
differential equations (we shall consider the gravitational
equations in
the Euclidean region)
\bea
\frac{da_1 (t)}{dt} &=& \frac{1}{(4\pi)^2} \left[ \xi (t) -
\frac{1}{6} \right]^2 \frac{N_s}{2}, \nn \\
\frac{da_2 (t)}{dt} &=& \frac{1}{120(4\pi)^2} \left( N_s+6N_f+
12N_A \right), \nn \\
\frac{da_3 (t)}{dt} &=& - \frac{1}{360(4\pi)^2} \left( N_s+11N_f+
62N_A \right),
\label{544}
\eea
where $N_s,N_f$ and $N_A$ are  the number of real
scalars, Dirac spinors and vectors, respectively (notice that the
running of $a_4(t)$ will not be meaningful for us, as we shall
see
below).

Owing to the fact that the theory under discussion is
multiplicatively
renormalizable, the effective Lagrangian satisfies the RG
equation
\beq
\left( \mu \frac{\partial}{\partial \mu} +
 \beta_i \frac{\partial}{\partial \lambda_i} -
 \gamma_i \phi_i \frac{\partial}{\partial \phi_i} \right)
L_{eff} (\mu,g_{\mu\nu}, \lambda_i, \phi_i )=0,
\label{545}
\eeq
where $\mu $ is the mass parameter, $\lambda_i =(g^2,h^2, f, \xi,
a_1,a_2,a_3,a_4)$ is the set of all coupling constants, the
$\beta_i$
are the corresponding $\beta$-functions and $\phi_i =(A_\mu,
\phi,
\psi)$.
The solution of Eqs. (\ref{545}) by the method of the
characteristics
gives (for all quantum fields we consider a zero background
field,
$\phi_i =0$):
\beq
L_{eff} (\mu,g_{\mu\nu}, \lambda_i)=
L_{eff} (\mu \, e^t,g_{\mu\nu}, \lambda_i (t)),
\label{546}
\eeq
where
\beq
\frac{d\lambda_i (t)}{dt} = \beta_i \left( \lambda_i (t)\right),
\ \ \ \
\ \ \lambda_i (0) =\lambda_i.
\label{547}
\eeq
Observe that for some of the coupling constants, the
corresponding
Eq. (\ref{547}) has been written above explicitly (Eq.
(\ref{544})),
while for a subset of them Eqs. (\ref{547}) have been actually
solved
(see Eqs. (\ref{542}) and (\ref{543})).

Actually, the idea itself of a
RG improvement procedure was suggested many years ago \cite{13}.
What we
do here is to make use once more of this interesting concept.
Physically, the meaning of expression (\ref{546}) is the
following:
$L_{eff}$ (called sometimes the Wilsonian effective action
\cite{5}) is
obtained through the above equations provided its functional form
for
some value of $t$ is known (usually it is the classical
Lagrangian that
serves as boundary condition at $t=0$). Another difficulty is
related
with the choice of $t$, which cannot be given a unique definition
due to
the presence, in general, of several different efective masses
(see the
discussions in Refs. \cite{6,7} concerning that point, for curved
and
for flat spaces, respectively).

There are different approaches to the gravitational effective
action
(for a general introduction, see \cite{8,1}). In the literature,
mainly
the case of a local effective action has been discussed (i.e.,
the
situation where the gravitational field is slowly varying).
One-loop
non-local effective actions have been considered in Refs.
\cite{9}-\cite{12} (see also the references therein), in
different
contexts, but almost exclusively the case of a free scalar field
theory
has been taken into account.

We will be interested in the situation where the gravitational
field is
weak, but rapidly varying, e.g.
\beq
\nabla \nabla R >> R^2.
\label{548}
\eeq
The non-local one-loop effective action for a free scalar field
theory
in this case has been calculated in Ref. \cite{10} (see also
\cite{11,12,14}), up to the second order on curvature
invariants.
Such a calculation is  quite tedious, moreover, its extension to
other
fields (especially, to interacting fields) is anything but
trivial (see
\cite{11} for a discussion and list of references).

We will make use of this RG improvement technique in our
calculation,
what is going to yield a correspondingly more precise result than
the
one that has been obtained till now by means of previous
approaches to
the problem. First, all those calculations have been carried out
in the
one-loop approximation, while ours here will yield the RG improved
effective
Lagrangian to leading-log order (through summation of all possible
logarithms) of
perturbation theory, i.e., clearly beyond one-loop. Secondly, the
theory
under discussion had been usually restricted to scalar fields,
while the
considerations here will be applicable to any renormalizable
theory on a
curved background, including the ordinary renormalizable models
of
quantum gravity, as R$^2$-gravity (see \cite{1} for a review). In
particular we will present results for an arbitrary
asymptotically free
GUT in curved spacetime (see \cite{1,2}).

To begin, using the general expression (\ref{546}) we can
write
explicitly the RG improved effective Lagrangian for the theory
(\ref{541}), employing the classical Lagrangian as boundary
condition:
\beq
L_{eff} = a_1(t) R^2+ a_2(t) C^2_{\mu\nu\alpha\beta} + a_3(t) G +
a_4(t)
\Box R,
\label{549}
\eeq
where the choice of RG parameter $t$ will be described below.
{}From the explicit one-loop calculation \cite{4,2}, the RG parameter
is found to be
\beq
t \sim {1 \over 2}\ln {-\Box + c_1 R \over \mu^2 },
\label{5410}
\eeq
where the constant $c_1$ is different in the different sectors
(scalar,
spinor and vector). By looking at (\ref{548}) one can see that
in order to
get the dominant contribution we may just keep the first term in
(\ref{5410}), i.e. $t\simeq (1/2) \ln (-\Box /\mu^2)$. From the
explicit
study of the non-local effective action \cite{10}-\cite{12} it
follows
that
the thing one has to understand is the way non-local form factors
act,
as formal operators obeying the variational rules of finite
matrices (in
the Lorentzian region). Note also that the terms $a_4(t) \Box R$
and
$a_3(t) G$ are still total derivatives after the RG improvement
(compare with the other regime in \cite{6} where these terms
become
important). Notice that a different way of understanding the appearance
of the $-\Box$ under the logarithm is to resort to RG considerations in
curved space \cite{1}, where we know that a scale transformations of
the metric, $g_{\mu\nu} \rightarrow e^{-2t} g_{\mu\nu}$, ought to be
performed. Since, under this transformation, $R^2 \rightarrow e^{4t}
R^2$ and $\Box \rightarrow e^{2t} \Box$, the  logarithm corresponding to
those terms becomes relevant in the high-energy limit $t\rightarrow
\infty$.

Finally, the RG improved non-local gravitational effective
Lagrangian
takes the form
\bea
L_{eff} &=& R \left\{ a_1 - \frac{(\xi - 1/6)^2 N_s}{2B^2g^2
(2b+1)}
\left[
\left( 1 + \frac{B^2g^2 \ln (- \Box /\mu^2)}{2(4\pi)^2}
\right)^{2b+1}
-1 \right] \right\} R \nn \\ &&+
 C_{\mu\nu\alpha\beta} \left[ a_2 + \frac{\ln (-\Box /\mu^2)}{240
(4\pi)^2} (N_s + 6N_f + 12N_A) \right]
 C^{\mu\nu\alpha\beta},
\label{5411}
\eea
where $a_1$ and $a_2$ are intital values for the corresponding
effective
couplings. Notice that with the above form factors the solutions
fulfill
the requeriment of asymptotic flateness \cite{14}. As it has been
discussed in Refs. \cite{14,15}, the coefficients of the terms
linear in  $\ln (-\Box)$ give a measure of the energy
radiation through the future null infinity.

Here we have obtained an effective Lagrangian, $L_{eff}$, which
sums
{\it all} the logarithms of perturbation theory, up to second
order
terms on curvature invariants on the background, of weak but
quickly
varying curvature. The theory under consideration is an
asymptotically
free GUT but, in principle, we can consider in the same way any
other
kind of renormalizable quantum field theory.

Notice, however, that the price one has to pay for the
universality of the approach (i.e., for the possibility to write
(\ref{5411}) for a variety of theories beyond the one-loop
approximation) is the fact that we cannot proceed to higher
orders in the curvature. The reason is  that the terms as $R^3$,
$R^4, \ldots$ are ultraviolet finite. At the same time, the
ordinary technique to one-loop order \cite{11,12} gives the
possibility, in principle, to calculate the non-local effective
action up to any desired order in the curvature ---although it is
quite complicated, already in the case of the scalar theory.
It turns out, therefore, that the two approaches complement each
other quite well.

Using $L_{eff}$ one can obtain the effective gravitational
equations. Adding the quantum matter-induced effective Lagrangian
(\ref{5411}) to the classical Einstein Lagrangian (without the
cosmological constant, for simplicity), one gets the effective
gravitational equations in close analogy with Refs.
\cite{11,12,14}. Before doing this, it is convenient to rewrite
\beq
 C_{\mu\nu\alpha\beta}^2 = G + 2 R_{\mu\nu}^2 - \frac{2}{3} R^2,
\label{5412}
\eeq
and to substitute it into Eq. (\ref{5411}). Then, one finds the
following Euclidean effective gravitational equations
\bea
&&- \frac{1}{8\pi G} \left( R_{\mu\nu} - \frac{1}{2} g_{\mu\nu} R
\right) + \left\{ a_1 + \frac{(\xi - 1/6)^2 N_s}{2B^2g^2 (2b+1)}
\left[
\left( 1 + \frac{B^2g^2 \ln (- \Box /\mu^2)}{2(4\pi)^2}
\right)^{2b+1}-1 \right] \right. \nn \\ &&-
\left. \frac{2}{3} a_2 - \frac{\ln (-\Box /\mu^2)}{360
(4\pi)^2} (N_s + 6N_f + 12N_A) \right\} \left[
4\nabla_\mu\nabla_\nu R - 4g_{\mu\nu} \Box R + {\cal O} (R^2)
\right] \label{5413} \\
&& + 2 \left[ a_2 + \frac{\ln (-\Box /\mu^2)}{240
(4\pi)^2} (N_s + 6N_f + 12N_A) \right]
\left[ 2\nabla_\mu\nabla_\nu R - g_{\mu\nu} \Box R -2 \Box
R_{\mu\nu}+ {\cal O} (R^2) \right] =0.   \nn
\eea
Observe that in order to obtain the effective gravitational
equations it is not necessary to take into account the
$g_{\mu\nu}$-dependence of the form factors. As was discussed in
Ref. \cite{14}, the effective gravitational equations can be used
in order to study the problem of collapse.

To be remarked is the fact that the above approach works well for
renormalizable models of quantum
gravity too. In order to exemplify this, let us consider
R$^2$-gravity under the form
\beq
L = \frac{1}{\lambda} \left( R_{\mu\nu} - \frac{1}{3} R^2 \right) -
\frac{\omega}{3\lambda} R^2.
\label{5414}
\eeq
Such a theory is multiplicatively renormalizable, being non-unitary in
the perturbative approach (for a general review and a list of
references, see \cite{1}). The RG improved non-local effective
Lagrangian corresponding to this theory, with the same gravitational
background (\ref{548}), can be easily constructed. The effective
gravitational equations are (for simplicity, only leading-log terms
have been kept)
\bea
&&\left[ \frac{1}{\lambda} +
\frac{133}{20 (4\pi)^2}  \ln (- \Box /\mu^2)
\right]
\left[
\nabla_\mu\nabla_\nu R - g_{\mu\nu} \Box R - 2 \Box R_{\mu\nu} + {\cal
O} (R^2) \right] \nn \\
&& + 2 \left[
- \frac{1}{3\lambda}
- \frac{\omega}{3\lambda}
 + \left( \frac{10}{9} \omega^2 + \frac{5}{3} \omega + \frac{5}{36}
\right) \frac{\ln (-\Box /\mu^2)}{2(4\pi)^2}
\right] \nn \\ && \hspace{6mm} \times \left[ 4\nabla_\mu\nabla_\nu R
-4g_{\mu\nu} \Box R + {\cal O} (R^2) \right] =0,
\label{5415}
\eea
where $\lambda$ and $\omega$ are the initial values for the corresponding
effective couplings.

Following now Refs. \cite{14,15} (it is explained there in which form
non-local effective action can be relevant for black-hole physics), we
can discuss the implications that the above non-local gravitational
action has concerning the flux of the vacuum radiation in an
asymptotically free GUT. Working with the asymptotically flat
(Lorentzian) solution of Eqs. (\ref{5413}) one may consider the
congruence $u(x)=$ const. of the light rays that can reach the future
null infinity $\cal{F}^+$. We shall denote, as in \cite{14}, by $r$  the
luminosity distance along rays and by $M(u)$  the Bondi mass at
 $\cal{F}^+$ (see \cite{17}). Then, the final expression for the
radiation corresponding to the vacuum energy in a spherically symmetric
state has been found to be the following \cite{14}:
\beq
\frac{d M(u)}{d u} = -\frac{1}{4\pi} (w_1+2w_2) \frac{d^2}{d^2u}
\int_{{\cal F}^-}^{{\cal F}^+} dr \, r \, R + {\cal O} (R^2),
\label{5416}
\eeq
where $w_1$ and $w_2$ are the coefficients of terms linear in $\ln
(-\Box)$ of (\ref{5411}), that is
\beq
L_{eff} = \left\{ R_{\mu\nu} \left[ a_1 - \frac{2}{3} a_2 -
\frac{w_1}{2(4\pi)^2} \ln \left( - \frac{\Box}{\mu^2} \right) \right]
R^{\mu\nu} + R \left[ 2a_2 -
\frac{w_2}{2(4\pi)^2} \ln \left( - \frac{\Box}{\mu^2} \right) \right] R
\right\},
\label{5417}
\eeq
where $a_1$ and $a_2$ can be taken to be zero and where $a_1(t)$ has
been expanded up to terms linear on $\ln (-\Box)$. Taking into account
the overall change of sign of $L_{eff}$ in the Lorentzian region, from
(\ref{544}) we obtain
\beq
w_1 = \frac{1}{60} \left( N_s + 6N_f + 12 N_A \right) , \ \ \ \
w_2 =-\frac{1}{180} \left( N_s + 6N_f + 12 N_A \right)+ \frac{N_s}{2}
\left( \xi - \frac{1}{6} \right).
\label{5418}
\eeq
In this way we can calculate the rate of the vacuum energy radiation
through the future null infinity, taking into account corrections to the
 GUT under consideration.
To be remarked is the fact that the choice of $\xi$ can influence this
rate of radiation significantly (\ref{5418}). Radiation disappears when the
null surface $u=$ const. comes very close to the horizon \cite{14,15}.
Then, in order to find the Hawking radiation \cite{16} one has to
calculate the next-to-leading correction in (\ref{5417}), namely the
${\cal O} (R^2)$-terms.

To summarize, using rather simple RG considerations, we have constructed
a RG improved non-local gravitational Lagrangian corresponding to a
general asymptotically free GUT and also to R$^2$-quantum gravity. The
corresponding effective gravitational equations have been written
down as well. It would be now of interest to study the applications of
these equations to black hole physics in more detail, since they
certainly modify a number of results obtained previously.

\vspace{5mm}


\noindent{\large \bf Acknowledgments}

We would like to thank Roberto Percacci and Sergio Zerbini for their
interest in this work.
SDO is grateful with the members of the Department ECM, Barcelona
University, for warm hospitality.
This work has been supported by DGICYT (Spain), project
Nos. PB93-0035 and SAB93-0024, by CIRIT (Generalitat de
Catalunya), and by RFFR, project 94-020324.


\newpage

\begin{thebibliography}{99}

\bibitem{8}
B.S. De Witt, {\it Dynamical Theory of Groups and Fields} (Gordon and
Breach, London, 1965).

\bibitem{1} I.L. Buchbinder, S.D. Odintsov and  I.L. Shapiro,
{\it Effective action in quantum gravity} (IOP, Bristol and
Philadelphia, 1992).

\bibitem{18}
G.W. Gibbons, S.W. Hawking and M.J. Perry,  Nucl. Phys. {\bf
B318} (1978) 141;
G.W. Gibbons and M.J. Perry,  Nucl. Phys. {\bf
B146} (1978) 90;
S.M. Christensen, M.J. Duff, G.W. Gibbons and M. Rocek,
Phys. Rev. Lett.  {\bf 45} (1980) 161;
I. Antoniadis, J. Iliopoulos and T.N. Tomaras,
Phys. Rev. Lett.  {\bf 56} (1986) 1319;
E.S. Fradkin and A.A. Tseytlin, Nucl. Phys. {\bf B234} (1984) 472;
S.D. Odintsov, Europhys. Lett. {\bf 10} (1989) 287;
T.R. Taylor and G. Veneziano, Nucl. Phys. {\bf B345} (1990) 210;
A.A. Bytsenko, S.D. Odintsov and S. Zerbini, Class. Quant. Grav. {\bf
12} (1995) 1.

\bibitem{9} R. Jordan, Phys. Rev. {\bf D33} (1986) 444;
L. Parker and D.J. Toms, Phys. Rev. {\bf D32}
(1985) 1409; E. Calzetta and B.L. Hu,
 Phys. Rev. {\bf D35} (1987) 495; J.P. Paz, Phys. Rev. {\bf D41} (1990)
1054.

\bibitem{10} I.G. Avramidi, Yad. Fiz. (Sov. J. Nucl. Phys.) {\bf 49}
(1989) 735.

\bibitem{11} G.A. Vilkovisky, Publ. Inst. Rech. Math. Advanc. R.C.P.
{\bf 25}, 43 (Strasbourg, 1992).

\bibitem{12} D. Dalvit and F. Mazzitelli, Phys. Rev. {\bf D50} (1994)
1001.

\bibitem{5} K.G. Wilson and J. Kogut,  Phys. Rep. {\bf 12}
(1974) 75.

\bibitem{3} N.P. Chang, A. Das and J. P\'erez-Mercader, Phys.
Rev. {\bf D22} (1980) 1429; {\bf D22} (1980) 1829.

\bibitem{4} B.L. Voronov and I.V. Tyutin,  Yad. Fiz. (Sov. J.
Nucl. Phys.) {\bf 23} (1976) 664.

\bibitem{2}
I.L. Buchbinder and S.D. Odintsov, Izw. VUZov. Fiz. (Sov. Phys.
J.) N12
(1983) 108; Yad. Fiz. (Sov. J. Nucl. Phys.) {\bf 40} (1984) 1338.

\bibitem{13} S. Coleman and E. Weinberg, Phys. Rev. {\bf D7} (1973)
1888.

\bibitem{6} E. Elizalde and S.D. Odintsov, Phys.
Lett. {\bf B321} (1994) 199;
E. Elizalde, K. Kirsten and S.D. Odintsov, Phys. Rev. {\bf D50}
(1994) 5137; E. Elizalde, S.D. Odintsov and A. Romeo, Phys. Rev. {\bf
D51} (1995) 1680.

\bibitem{7}
C. Ford, D.R.T. Jones, P.W. Stephenson and M. Einhorn,
 Nucl. Phys.  {\bf B395} (1993) 405.

\bibitem{14} A.G. Mirzabekian and G.A. Vilkovisky, Phys. Lett. {\bf
B317} (1993) 517.

\bibitem{15} V.P. Frolov and G.A. Vilkovisky, Phys. Lett. {\bf
B106} (1981) 307.

\bibitem{17}
R.M. Wald, {\it General Relativity} (Chicago, 1984).

\bibitem{16}
S.W. Hawking,  Commun. Math. Phys. {\bf D37} (1975) 199.


\end{thebibliography}
\end{document}